\documentclass[onecolumn,aps,preprintnumbers,prb,superscriptaddress]{revtex4}%
\usepackage{graphicx}
\usepackage{dcolumn}
\usepackage{bm}
\usepackage{times}
\usepackage[colorlinks=true,linkcolor=blue,anchorcolor=blue,citecolor=blue,urlcolor=blue]%
{hyperref}
\usepackage{multirow}
\usepackage{amsmath}
\usepackage{amsfonts}
\usepackage{amssymb}
\usepackage{color}%
\begin{document}
\title{Topological orbital superfluid with chiral $d$-wave order in a rotating
optical lattice}
\author{Ningning Hao}
\email{haon@hmfl.ac.cn}
\affiliation{Anhui Province Key Laboratory of Condensed Matter Physics at Extreme
Conditions, High Magnetic Field Laboratory, Chinese Academy of Sciences, Hefei
230031, Anhui, China}
\author{Huaiming Guo}
\affiliation{Department of Physics, Beihang University, Beijing 100191, China}
\author{Ping Zhang}
\email{zhang\_ping@iapcm.ac.cn}
\affiliation{Institute of Applied Physics and Computational Mathematics, Beijing 100088, China}
\affiliation{Beijing Computational Science Research Center, Beijing 100193, China}

\begin{abstract}
Topological superfluid is an exotic state of quantum matter that possesses a
nodeless superfluid gap in the bulk and Andreev edge modes at the boundary of
a finite system. Here, we study a multi-orbital superfluid driven by
attractive $s$-wave interaction in a rotating optical lattice. Interestingly,
we find that the rotation induces the inter-orbital hybridization and drives
the system into topological orbital superfluid in accordance with
intrinsically chiral $d$-wave pairing characteristics. Thanks to the
conservation of spin, the topological orbital superfluid supports four rather
than two chiral Andreev edge modes at the boundary of the lattice. Moreover,
we find that the intrinsic harmonic confining potential forms a circular
spatial barrier which accumulates atoms and supports a mass current under
injection of small angular momentum as external driving force. This feature
provides an experimentally detectable phenomenon to verify the topological
orbital superfluid with chiral $d$-wave order in a rotating optical lattice.

\end{abstract}
\maketitle

%\author{Huaiming Guo}
%\affiliation{Department of Physics, Beihang University, Beijing, 100191, China}
%\affiliation{Department of Physics, The University of Hong Kong, Pokfulam Road, Hong Kong, China}
%\author{Shun-Qing Shen}
%\affiliation{Department of Physics, The University of Hong Kong, Pokfulam Road, Hong Kong, China}

%\pacs{67.85.-d, 03.75.Ss, 74.20.Fg, 74.20.-z}

\section{Introduction}

Orbital degrees of freedom play a significant role to produce various exotic
quantum states in complex condensed-matter systems, such as high temperature
superconductors and quantum magnetic insulators. Recent experimental
realizations of multi-orbital systems with ultra-cold
atoms\cite{Muller2007PRL,Wirth2011NP,Soltan2012NP,Parker3013NP} have promoted
the theoretical studies of high orbital physics in optical lattices, where a
series of exotic quantum states have been
proposed\cite{Sun2012NP,Li2013NC,Liu2014NC,Liu2015Arxiv,Wu2008PRL,Wu2008PRL-2,Zhai2013PRA}%
. Among them, one of remarkable characteristics is that the orbital
hybridization can play the same role as spin-orbital coupling or artificial
gauge fields which are the key ingredient to drive topologically insulating or
superconducting states\cite{Qi2011RMP,Hasan2010RMP}. Therefore, topologically
nontrivial many-body states can be implemented in multi-orbital systems in the
absence of spin-orbital couplings. There exist several methods to induce the
orbital hybridization in the context of cold atom systems, including many-body
interaction effect\cite{Sun2012NP}, lattice
shaking\cite{Struck2011SC,Hauke2012PRL,Parker2013NP,Koghee2012PRA}, and local
rotation\cite{GemelkeARXIV2010}. The relevant quantum states including
topological semimetal\cite{Sun2012NP} and topological band
insulators\cite{Hauke2012PRL,Zhang2014PRA,Zheng2014PRA,Wu2008PRL-2} have been proposed.

Recently, the superfluid of bosons with chiral odd-frequency orders, i.e.,
$p+ip$-wave and $f+if$-wave, have been experimentally realized in
multi-orbital cold-atom
systems\cite{Wirth2011NP,Matthias2011PRL,Matthias2013NJP}. For the fermions,
however, it is still a big challenge to realize the superfluid states with
chiral odd-frequency orders, because the atom loss is strong near the Feshbach
resonance in high-frequency channels\cite{Chin2010RMP}. Theoretically, thanks
to the Rashba spin-orbital couplings, the topological superfluids of fermions
with chiral odd-frequency orders have been proposed to emerge in $s$-wave
channel of the Feshbach resonance.
\cite{Zhang2009PRL,Sato2009PRL,Liu2012PRA,Hao2013PRA}. In comparison with
well-studied chiral odd-frequency superfluids of fermions, the superfluids of
fermions with chiral even-frequency orders are rarely studied, and only some
candidate materials are proposed to have the chiral even-frequency orders due
to the unconventional superconducting pairing in condensed-matter
systems\cite{Krishana1997Science,Laughlin1998PRL,Nandkishore2012NP,Liu2013PRL}%
. More recently, a checkerboard lattice in a periodic Floquet driving field
was proposed to support the chiral $d$-wave superfluid, where the sublattice
degrees of freedom plays a key role and the periodic Floquet driving field
induces the hybridization of two sublattices\cite{Zhang2015Arxiv}. In this
paper, we propose that a superfluid state of fermions with a chiral $d$-wave
order can be implemented in a rotating multi-orbital optical lattice. In our
proposal, the key ingredients to drive the underlying nontrivial topology of
the multi-orbital superfluid state with a chiral $d$-wave order come from the
two orbitals that are the counterparts of spin degrees of freedom in
spin-orbital coupling, and the inter-orbital hybridization is induced by the
local rotation with same frequency for every individual lattice site, which
can be experimentally realized \cite{GemelkeARXIV2010}. Interestingly,
different from conventional chiral $d$-wave topological superfluid which
supports two chiral Andreev edge modes at the boundary of the system, the
topological orbital superfluid here supports four chiral Andreev edge modes
due to the conservation of spin. More importantly, we find that the spatial
barrier structure spontaneously formed by the intrinsic harmonic confining
potential separates the trivial and nontrivial superfluid states, accumulates
cold atoms and supports a mass current under injection of small angular
momentum as the external driving force. These features can be experimentally
adopted to verify the topologically non-trivial superfluid states. In
comparison with the chiral $p$-wave and $f$-wave topological superconductor
and
superfluid\cite{Read2000PRB,Fu2008PRL,Sau2010PRL,Zhang2009PRL,Sato2009PRL,Mao2011PRL,Hao2013PRA,Wang2012NJP,
Hao2010PRB,Shen2012Book}, where the spin-orbital couplings are essential, the
chiral $d$-wave topological superfluid here only requires the orbital
hybridization. Therefore, our proposal provides a possible route to explore
topological superfluids with chiral $d$-wave order in multi-orbital cold-atom systems.

The paper is organized as follows. In section II, we discuss the
implementation of the multi-orbital system with a specific configuration of
laser beams, and construct the effective Hamiltonian to describe the
multi-orbital system. In section III, we study the homogeneous superfluid
state with self-consistent mean-field approximation, and discuss the
topological properties of the homogeneous superfluid state. In section IV, we
discuss the inhomogeneous superfluid state modulated by the harmonic confining
potential. In section V, we discuss the experimental scheme and present a
brief summary.

\section{Optical lattice and Model}

\begin{figure}[pt]
\begin{center}
\includegraphics[width=1.0\linewidth]{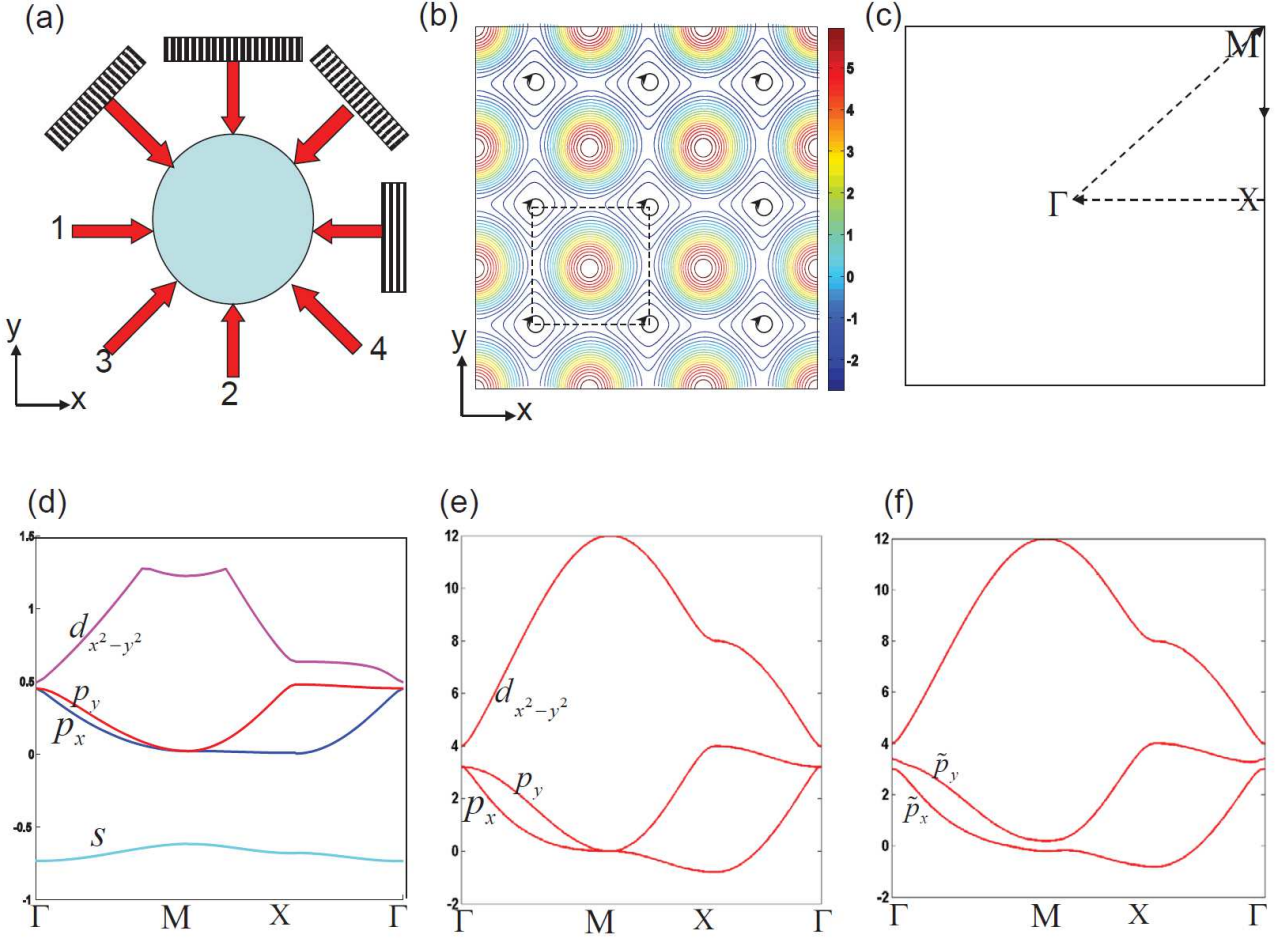}
\end{center}
\caption{(Color Online) (a) Four retro-reflected laser beams are adopted to
create the lattice potential in Eq. (1). (b) The contour of the lattice
potential forms a two-dimensional optical lattice, and the atoms are trapped
at the minima of the potential. The small circle with arrow at each minimum
represents the on-site rotation. Here, $V_{1}=1.2E_{R}$ and $V_{2}=0.4E_{R}$
(c) The Brillouin zone and high-symmetry points. (d) The single-particle
energy spectrum along high-symmetry lines in the unit of $E_{R}$ for the four
lowest bands through plane wave expansion calculation about the lattice
potential. (e) and (f) The single-particle energy spectrum along high-symmetry
lines from the tight-binding Hamiltonian in Eq.(\ref{Ham_tb}) without and with
on-site rotation.
To guarantee the consistence of the energy scales between the bands from plane wave expansion
calculation about the lattice potential in (d) and the bands from tight-binding calculations in (e) and (f), the energy is measured in the unit of $t_{pp}$ with $t_{pp}=0.1E_{R}$ in (e) and (f). Other parameters are $t_{dd}=1$, $t_{pd}=1$, $t_{pp}^{\prime}=0.2$, $\delta=6.4$, $\mu_{0}=-1.6$, $V_{t}=0$ and
$h\Omega_{z}=0$ in (e) and $h\Omega_{z}=0.2$ in (f). }%
\label{fig_laser}%
\end{figure}

We consider a balanced mixture of fermion atoms with two internal states
labeled by the spin index $\sigma$. The atoms are loaded in an
isotropic 2D square optical lattices. To introduce the couplings between
different p orbital bands, one effective approach is to rotate the optical
lattice with same rotation frequency $\Omega_{z}$ for every individual lattice
site\cite{GemelkeARXIV2010}. An alternative approach would be to directly
couple the states with a drive laser\cite{Pinheiro2013PRL}. Finally, the
trapped atoms are turned close to a Feshbach resonance to produce attractive
$s$-wave interactions. The lattice potential takes the form,%
\begin{equation}
V(x,y)=V_{1}[\cos k_{L}x+\cos k_{L}y]+2V_{2}\cos k_{L}x\cos k_{L}y.
\label{potentialVxy}%
\end{equation}
Here, $V_{1}$ and $V_{2}$ are the optical lattice potentials and $k_{L}$ is
the wave-vector of laser fields. The realization of lattice potential $V(x,y)$
in Eq. (\ref{potentialVxy}) has been proposed for the case $V_{2}/V_{1}%
>1/2$\cite{Sun2012NP}. Here, we consider the case $V_{2}/V_{1}<1/2$, and the
configuration of optical lattices under the condition $V_{2}/V_{1}<1/2$ can be
implemented through four retro-reflected laser beams as shown in Fig.
\ref{fig_laser}(a). The electric field generated by each laser beam is%

\begin{equation}
\vec{E}_{j}(\vec{r},t)=E_{j,0}\vec{e}_{j}\cos(\vec{k}_{j}\cdot\vec{r}%
_{j})e^{-i(\omega_{j}t+\varphi_{j})}, \label{electric_field}%
\end{equation}
where $\vec{e}_{j}$, $\omega_{j}$, and $\varphi_{j}$ are the polarization
vector, the frequency, and the phase of the laser field, respectively. The
parameters for each laser beams are summarized in Table \ref{laser-field}. The
corresponding light-shift potential is\begin{table}[pt]
\caption{The parameters of the electric fields of four laser beams shown in
Fig. 1(a).}%
\label{laser-field}
\begin{tabular}
[c]{llllll}\hline\hline
$j$ & $E_{j0}$ & $\vec{e}_{j}$ & $\vec{k}_{j}$ & $\omega_{j}$ & $\varphi_{j}%
$\\\hline
$1$ & $\epsilon_{1}$ & $(1,0,0)$ & $(k_{L}/2,0)$ & $\omega_{0}$ & $0$\\
$2$ & $\epsilon_{1}$ & $(0,1,0)$ & $(0,k_{L}/2)$ & $\omega_{0}$ & $0$\\
$3$ & $\epsilon_{2}$ & $(0,0,1)$ & $(k_{L}/2,k_{L}/2)$ & $\omega_{0}$ & $0$\\
$4$ & $\epsilon_{2}$ & $(0,0,1)$ & $(k_{L}/2,-k_{L}/2)$ & $\omega_{0}$ &
$0$\\\hline
\end{tabular}
\end{table}%
\begin{equation}
V(x,y)=-\chi|\sum_{j}\vec{E}_{j}(\vec{r},t)|^{2}, \label{lattice_potential}%
\end{equation}
with $\chi$ denoting the real part of the polarizability. By adopting the
parameters in Table \ref{laser-field}, we can get the lattice potential shown
in Eq.(\ref{potentialVxy}) with an irrelevant constant shift $V_{0}%
=-\chi(\epsilon_{1}^{2}+\epsilon_{2}^{2})$. Here, $V_{1}=-\chi(\epsilon
_{1}^{2}/2+\epsilon_{2}^{2})$, and $V_{2}=-\chi\epsilon_{2}^{2}/2$. The
condition $V_{2}/V_{1}<1/2$ can be achieved for arbitrary nonzero
$\epsilon_{1}$ and $\epsilon_{2}$ and blue detuning with $\chi<0$. Here, we
set $V_{1}=1.2E_{R}$ and $V_{2}=0.4E_{R}$. $E_{R}=\frac{h^{2}}{2ma^{2}}$ is
the recoil energy and $a$ is the lattice constant.

The contour of $V(x,y)$ is shown in Fig. \ref{fig_laser}(b). The lowest four
band structures from the plane-wave expansion approximation upon the potential
$V(x,y)$ in Eq. (\ref{potentialVxy}) are shown in Fig. \ref{fig_laser} (d). It
is straightforward to check that the splitting between two middle $p_{x}$ and
$p_{y}$ bands off the high-symmetry point are induced by the coupling to the
higher $d_{x^{2}-y^{2}}$ band\cite{Sun2012NP}. Consider the three orbitals of
$p_{x}$, $p_{y}$ and $d_{x^{2}-y^{2}}$ shown in Fig. \ref{fig_laser}(e), a
tight-binding (TB) Hamiltonian can be constructed to described the band
structures of the fermionic square lattice, i.e.,%
\begin{equation}
H_{tb}=H_{d}+H_{p}+H_{dp}, \label{Ham_tb}%
\end{equation}
with%
\begin{equation}
H_{d}=\sum_{<i,j>\sigma}[-t_{dd}+(\delta-\mu_{i})\delta_{ij}]d_{i,\sigma
}^{\dag}d_{j,\sigma}, \label{Ham_tb1}%
\end{equation}%
\begin{align}
H_{p}  &  =-\sum_{i\sigma,l=x,y}\mu_{i}p_{l,i,\sigma}^{\dag}p_{l,i,\sigma
}+ih\Omega_{z}\sum_{i\sigma}p_{x,i,\sigma}^{\dag}p_{y,i,\sigma}%
+H.c.\nonumber\\
&  +t_{pp}\sum_{i\sigma,l=x,y}p_{l,i,\sigma}^{\dag}p_{l,i+e_{l},\sigma
}+H.c.\nonumber\\
&  -t_{pp}^{\prime}\sum_{i\sigma,l=x,y,\bar{l}=-l}p_{l,i,\sigma}^{\dag
}p_{l,i+e_{\bar{l}},\sigma}+H.c., \label{Ham_tb2}%
\end{align}%
\begin{equation}
H_{dp}=t_{pd}\sum_{i\sigma,l=x,y}[p_{l,i+e_{l},\sigma}^{\dag}d_{i,\sigma
}-p_{l,i-e_{l},\sigma}^{\dag}d_{i,\sigma}]+H.c. \label{Ham_tb3}%
\end{equation}
Here, $\mu_{i}=\mu_{0}+V_{trap}(i_{x},i_{y})$ with
\begin{equation}
V_{trap}(i_{x},i_{y})=V_{t}[(i_{x}-\frac{N_{x}+1}{2})^{2}+(i_{y}-\frac
{N_{y}+1}{2})^{2}] \label{Vtrap}%
\end{equation}
being the weak harmonic confining potential to stabilize the optical lattice.
$p_{x/y,i,\sigma}^{\dag}$, and $d_{i,\sigma}^{\dag}$ are the fermion creation
operators for atoms in the relevant $p_{x}$, $p_{y}$ and $d_{x^{2}-y^{2}}$
orbitals. We first set $V_{t}=0$ to simplify the discussions and recover it
later.
Note that all the energy scales are measured in the unit of $t_{pp}$ as explained in the caption of Fig. \ref{fig_laser} in the following parts of the paper if not special specified.
The energy spectra of TB Hamiltonian in Eq. (\ref{Ham_tb}) are shown in Figs.
\ref{fig_laser}(e) and \ref{fig_laser}(f). It can be found that the TB
Hamiltonian in Eq. (\ref{Ham_tb}) gives a good description of the band
structures of lattice potential, and the on-site rotation in the second term
in Eq. (\ref{Ham_tb2}) induces the orbital hybridization to break the
degeneracy of $p_{x}$ and $p_{y}$ bands around the $\Gamma$ and $M$ points
shown in Fig. \ref{fig_laser}(c).

When the fermion atoms are loaded into the two $p_{x}$ and $p_{y}$ bands, the
attractive $s$-wave interactions from the Feshbach resonance give the
two-orbital attractive Hubbard interactions as follows\cite{Zhang2010PRA},%
\begin{align}
H_{int}  &  =U\sum_{il}n_{il\uparrow}n_{il\downarrow}-\frac{J}{2}\sum
_{i}[2\mathbf{S}_{ix}\cdot\mathbf{S}_{iy}+\frac{1}{2}n_{ix}n_{iy}]\nonumber\\
&  +\frac{J}{2}\sum_{i}n_{ix}n_{iy}+J_{\Delta}\sum_{i}p_{ix\uparrow}^{\dag
}p_{ix\downarrow}^{\dag}p_{iy\downarrow}p_{iy\uparrow}+H.c. \label{H_int}%
\end{align}
Here, the first term is the intra-orbital attractive interaction, and the
second term is the Hund's coupling with the spin operator $\mathbf{S}%
_{il}=\frac{1}{2}p_{il,\alpha}^{\dag}\sigma_{\alpha\beta}p_{il,\beta}$ and
$l=x,y$. $U$ and $J$ take the following forms,%

\begin{align}
U  &  =4\pi\hbar^{2}a_{s}/m\int dr|\omega_{x/y}(r)|^{4},\label{IntU}\\
J  &  =4\pi\hbar^{2}a_{s}/m\int dr|\omega_{x}(r)|^{2}|\omega_{y}(r)|^{2}.
\label{IntJ}%
\end{align}
Here, $a_{s}$ is the $s$-wave scattering length with negative value, i.e.,
$a_{s}<0$. $\omega_{x/y}(r)$ are the Wannier functions of $p_{x/y}$ orbitals.
The third term in Eq. (9) is the inter-orbital attractive interaction with
$n_{il}=n_{il,\uparrow}+n_{il,\downarrow}$. The fourth term is the pair
hopping term. Furthermore, we have $J=2U/3$ and $J_{\Delta}=U/3$%
\cite{Zhang2010PRA}. Note that the Hund's coupling and inter-orbital
interaction have same amplitudes, which are different from the electron
system.
The interaction terms shown in Eqs.(9)-(11) are obtained under the harmonic approximation. It is shown that the an-harmonicity of the optical lattice can affect the properties of the multi-orbital system\cite{Collin2010PRA,Tomasz2013PRL}. In particular, the intra-orbital interaction $U_{xx}$ is not equal to $U_{yy}$, and the inter-orbital interaction $J$ is off $2U_{xx}/3$. Such imbalance can induce the modulations of superfluid order parameters. However, the topological superfluid is robust against such small modulations, because nontrivial topology is the global feature of superfluid. For simplification, we neglect the irrelevant an-harmonic effects in the present work.%

\section{Homogeneous superfluid states with chiral d-wave order}

Now, we turn to consider the homogeneous superfluid state with $V_{t}=0$ in
Eq. (\ref{Vtrap}) and the superfluid state is driven by the attractive
interaction in Eq. (\ref{H_int}). The spin-singlet superfluid pairing
operators are defined as%
\begin{equation}
\hat{\Delta}_{s,ll^{\prime}}(k)=\sum_{\sigma\sigma^{\prime}}\frac{[i\sigma
_{y}]_{\sigma\sigma^{\prime}}}{4}[p_{l,k\sigma}p_{l^{\prime},-k\sigma^{\prime
}}+p_{l^{\prime},k\sigma}p_{l,-k\sigma^{\prime}}]. \label{op_s}%
\end{equation}
Then, we have%
\begin{align}
H_{int}  &  =U\sum_{l}\hat{\Delta}_{s,ll}^{\dag}\hat{\Delta}_{s,ll}+J_{\Delta
}\sum_{l\neq l^{\prime}}\hat{\Delta}_{s,ll}^{\dag}\hat{\Delta}_{s,l^{\prime
}l^{\prime}}\nonumber\\
&  +2J\sum_{l>l^{\prime}}\hat{\Delta}_{s,ll^{\prime}}^{\dag}\hat{\Delta
}_{s,ll^{\prime}} \label{Hint1}%
\end{align}
with
\begin{equation}
\hat{\Delta}_{s,ll^{\prime}}=\sum_{k}\hat{\Delta}_{s,l^{\prime}l^{\prime}}(k).
\label{deltat}%
\end{equation}
Note that the spin-triplet pairing parts disappear, because the Hund's
coupling and inter-orbital interaction have the same amplitudes. Through the
mean-field approximation, $\Delta_{s,ll^{\prime}}=\langle\hat{\Delta
}_{s,ll^{\prime}}\rangle$, $H_{int}$ can be decoupled to be
\begin{align}
H_{int}^{p}  &  =\sum_{l,k}(U\Delta_{s,ll}+J_{\Delta}\Delta_{s,\bar{l}\bar{l}%
})\hat{\Delta}_{s,ll}^{\dag}(k)+H.c.\nonumber\\
&  +\sum_{k}2J\Delta_{s,xy}\hat{\Delta}_{s,xy}^{\dag}(k)+H.c.+h_{con}
\label{Hintmf}%
\end{align}
with
\begin{equation}
h_{con}=-U\sum_{l}|\Delta_{s,ll}|^{2}-2J|\Delta_{s,xy}|^{2}-2J_{\Delta
}\operatorname{Re}(\Delta_{s,xx}\Delta_{s,yy}^{\ast}). \label{hcon}%
\end{equation}
The homogeneous superfluid state can be described by the mean-field
Hamiltonian in the Nambu basis: $\Psi(k)=[d_{k\uparrow},p_{x,k\uparrow
},p_{y,k\uparrow},d_{k\downarrow}^{\dag},p_{x,k\downarrow}^{\dag
},p_{y,k\downarrow}^{\dag},d_{k\downarrow},p_{x,k\downarrow},p_{y,k\downarrow
},d_{k\uparrow}^{\dag},p_{x,k\uparrow}^{\dag},p_{y,k\uparrow}^{\dag}]^{t}$,%
\begin{equation}
H_{mf}=\sum_{k}\frac{1}{2}\Psi^{\dag}(k)\left[
\begin{array}
[c]{cccc}%
H_{tb}(k) & \mathbf{\Delta} &  & \\
\mathbf{\Delta}^{\dag} & -H_{tb}^{\ast}(-k) &  & \\
&  & H_{tb}(k) & -\mathbf{\Delta}\\
&  & -\mathbf{\Delta}^{\dag} & -H_{tb}^{\ast}(-k)
\end{array}
\right]  \Psi(k)+C. \label{Hmf}%
\end{equation}
Here, $C$ is an operator-independent constant term. $\mathbf{\Delta}$ is a
$3\times3$ matrix and takes the following form,%
\begin{equation}
\mathbf{\Delta}=\left[
\begin{array}
[c]{ccc}%
0 & 0 & 0\\
0 & U\Delta_{s,xx}+J_{\Delta}\Delta_{s,yy} & 2J\Delta_{s,xy}\\
0 & 2J\Delta_{s,xy} & U\Delta_{s,yy}+J_{\Delta}\Delta_{s,xx}%
\end{array}
\right]  . \label{order_parameter1}%
\end{equation}
\begin{figure}[ptb]
\begin{center}
\includegraphics[width=1.0\linewidth]{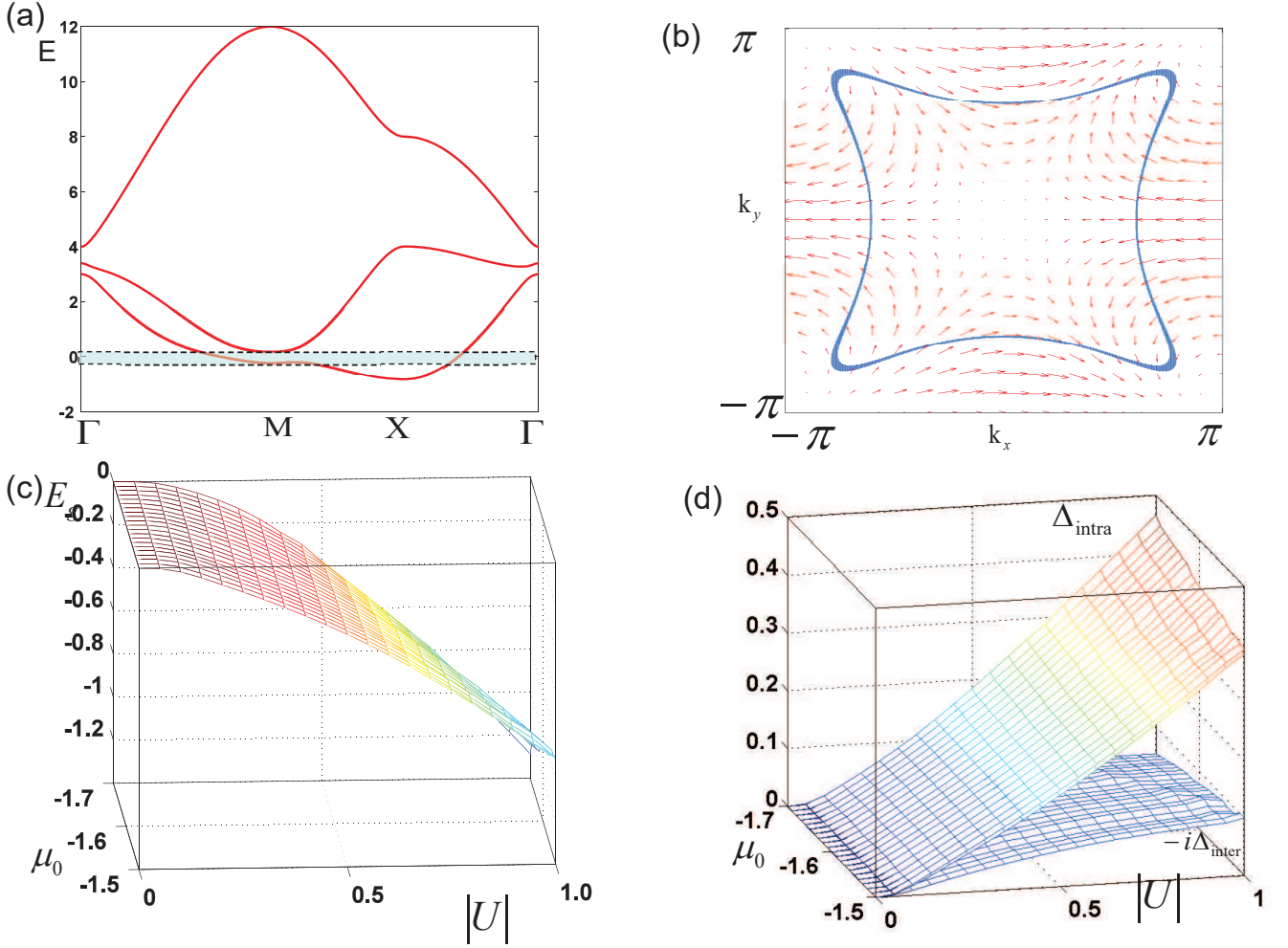}
\end{center}
\caption{(Color Online) (a) The band structure along the high-symmetry lines,
and the filling is in the shadowed regime by tuning the chemical potential.
(b) The closed curve denotes the single Fermi surface. The red arrows denote
the vector field of [$\xi_{-}(k)$, $\xi_{xy}(k)$]. Here, the parameters are
same as these in Fig. 1(f). (c) The zero-temperature ground-state energy of
superfluid state as change as chemical potential $\mu_{0}$ and interaction
amplitude $|U|$. (d) The intra-and inter-orbital superfluid order parameters
as change as chemical potential $\mu_{0}$ and interaction amplitude $|U|$.
Here, $\Delta_{intra}=\Delta_{22}$ with $\Delta_{33}=\Delta_{22}$, and
$\Delta_{inter}=\Delta_{23}$. The explicit expressions of $\Delta_{22}$,
$\Delta_{23}$ and $\Delta_{33}$ are shown in Eq.(\ref{order_parameter1}),
which are the relevant matrix elements. The mesh of $k_{x}\times
k_{y}=51\times51$.}%
\label{fig_op}%
\end{figure}The mean-field Hamiltonian in Eq. (\ref{Hmf}) can be
self-consistently solved with respect to the minimum of ground state energy,
i.e.,
\begin{equation}
E_{g}=h_{con}-\frac{1}{4\pi^{2}}\sum_{n=1}^{3}\int d^{2}\mathbf{k[}%
|E_{n}^{(s)}(k)|-|E_{n}^{(0)}(k)|], \label{Eg}%
\end{equation}
where, $E_{n}^{(s)}(k)$ and $E_{n}^{(0)}(k)$ are the eigen-energy spectra of
the superfluid state and normal state. Here, we focus on the filling lying in
the band splitting around the $M$ point induced by the orbital hybridization
as shown in Fig. \ref{fig_op}(a). The typical Fermi surface is shown in Fig.
\ref{fig_op}(b). From Eq. (\ref{order_parameter1}), we can find
that the superfluid order parameter in the intra-$p_{x}$ orbital channel is
$\Delta_{22}=U\Delta_{s,xx}+J_{\Delta}\Delta_{s,yy}$ while the superfluid
order parameter in the intra-$p_{y}$ orbital channel is $\Delta_{33}%
=U\Delta_{s,yy}+J_{\Delta}\Delta_{s,xx}$. To maximize the superfluid gap, one
can find that $\Delta_{s,xx}\Delta_{s,yy}>0$ is favorable to obtain the
largest amplitudes of $\Delta_{22}$ and $\Delta_{33}$. The numerical results
for the ground state energy and superfluid order parameters as functions of
chemical potential $\mu_{0}$ and interaction amplitude $|U|$ are shown in
Figs. \ref{fig_op}(c) and \ref{fig_op}(d), from which the intra-orbital
$\Delta_{22}$ and $\Delta_{33}$ are degenerate in the whole parameter regime.
It means that $\Delta_{22}=\Delta_{33}$, and the only choice is $\Delta
_{s,xx}\Delta_{s,yy}>0$ thanks to $UJ_{\Delta}>0$. The aforementioned analyses
are consistent, and one can achieve that the superfluid ground states favor
$\Delta_{s,xx}$ and $\Delta_{s,yy}$ with same sign to maximize the superfluid
gap and to minimize the ground state energy. Furthermore, we can find that
the inter-orbital $\Delta_{23}$, which is also the matrix element in Eq.
(\ref{order_parameter1}), is purely imaginary, and much smaller than
$\Delta_{22/33}$. The reason lies in that the inter-orbital $\Delta_{23}$ is
induced by the orbital hybridization and modulated by $\Omega_{z}$. It is
conceivable that the strength of inter-orbital $\Delta_{23}$ could be
comparable to intra-orbital $\Delta_{22/33}$ when $\Omega_{z}$ is large
enough. However, the $\Delta_{23}$ has no relation with the topological nature
of the superfluid state, we only focus on the case with $\Omega_{z}$ set here.

In order to reveal the underlying topological nature of the superfluid states,
we first investigate the band characteristics of the normal states. As shown
in Fig. \ref{fig_laser}(e), the full separation between the $d$ band and $p$
bands guarantees the feasibility to downfold the Hamiltonian from the space
spanned by $d$ and $p$ orbitals to the space spanned by two effective
$\tilde{p}$ orbitals shown in Fig. \ref{fig_laser}(f). When $V_{t}=0$, the
translation symmetry allows ones to write the TB Hamiltonian in momentum space
under the effective basis $\tilde{\psi}_{\sigma}(k)=[\tilde{p}_{x,k,\sigma
},\tilde{p}_{y,k,\sigma}]^{t}$, i.e.,
\begin{equation}
\tilde{H}_{tb}=\sum_{k\sigma}\tilde{\psi}_{\sigma}^{\dag}(k)\tilde{H}%
_{tb}(k)\tilde{\psi}_{\sigma}(k). \label{Htb2}%
\end{equation}
Here,
\begin{equation}
\tilde{H}_{tb}(k)=\frac{1}{2}\xi_{+}(k)-\mu_{0}+\xi_{xy}(k)\sigma_{x}%
-h\Omega_{z}\sigma_{y}+\frac{1}{2}\xi_{-}(k)\sigma_{z}, \label{Htb30}%
\end{equation}
and%
\begin{align}
\xi_{\pm}(k)  &  =2(\tilde{t}_{pp}\mp\tilde{t}_{pp}^{\prime})(\cos k_{x}%
\pm\cos k_{y}),\label{kesa1}\\
\xi_{xy}(k)  &  =4\tilde{t}_{xy}\sin k_{x}\sin k_{y}. \label{kesa2}%
\end{align}
The Pauli matrices $\sigma_{i}$ with $i=x,y,z$ span the two effective
$\tilde{p}_{x}$ and $\tilde{p}_{y}$ orbital space. The effective TB
Hamiltonian $\tilde{H}_{tb}$ can be rewritten in the basis spanned by the
orbital angular momentum eigen-state, i.e.,
\begin{equation}
\bar{H}_{tb}=\sum_{k\sigma}\bar{\psi}_{\sigma}^{\dag}(k)\bar{H}_{tb}%
(k)\bar{\psi}_{\sigma}(k). \label{Htb3}%
\end{equation}
Here, $\bar{\psi}_{\sigma}^{\dag}(k)=[\bar{p}_{+,k,\sigma}^{\dag},\bar
{p}_{-,k,\sigma}^{\dag}]$ with $\bar{p}_{\pm,k\sigma}^{\dag}=\frac{1}{\sqrt
{2}}[\tilde{p}_{x,k\sigma}^{\dag}\pm i\tilde{p}_{y,k\sigma}^{\dag}]$, and%

\begin{equation}
\bar{H}_{tb}(k)=\frac{1}{2}\xi_{+}(k)-\mu_{0}+\frac{1}{2}\xi_{-}(k)s_{x}%
+\xi_{xy}(k)s_{y}-h\Omega_{z}s_{z}. \label{Htb4}%
\end{equation}
The Pauli matrices $s_{i}$ with $i=x,y,z$ span the two effective $\bar{p}_{+}$
and $\bar{p}_{-}$ orbital space. In the absence of $\Omega_{z}$,
$[F_{x}(k)=\frac{1}{2}\xi_{-}(k),F_{y}(k)=\xi_{xy}(k)]$ forms a vector field
in momentum space shown in Fig. \ref{fig_op}(b). Then, the band degeneracy
point at the $M$ point can be mapped into a vortex in the momentum space with
integer winding number\cite{Sun2012NP}, i.e.,\begin{figure}[ptb]
\begin{center}
\includegraphics[width=1.0\linewidth]{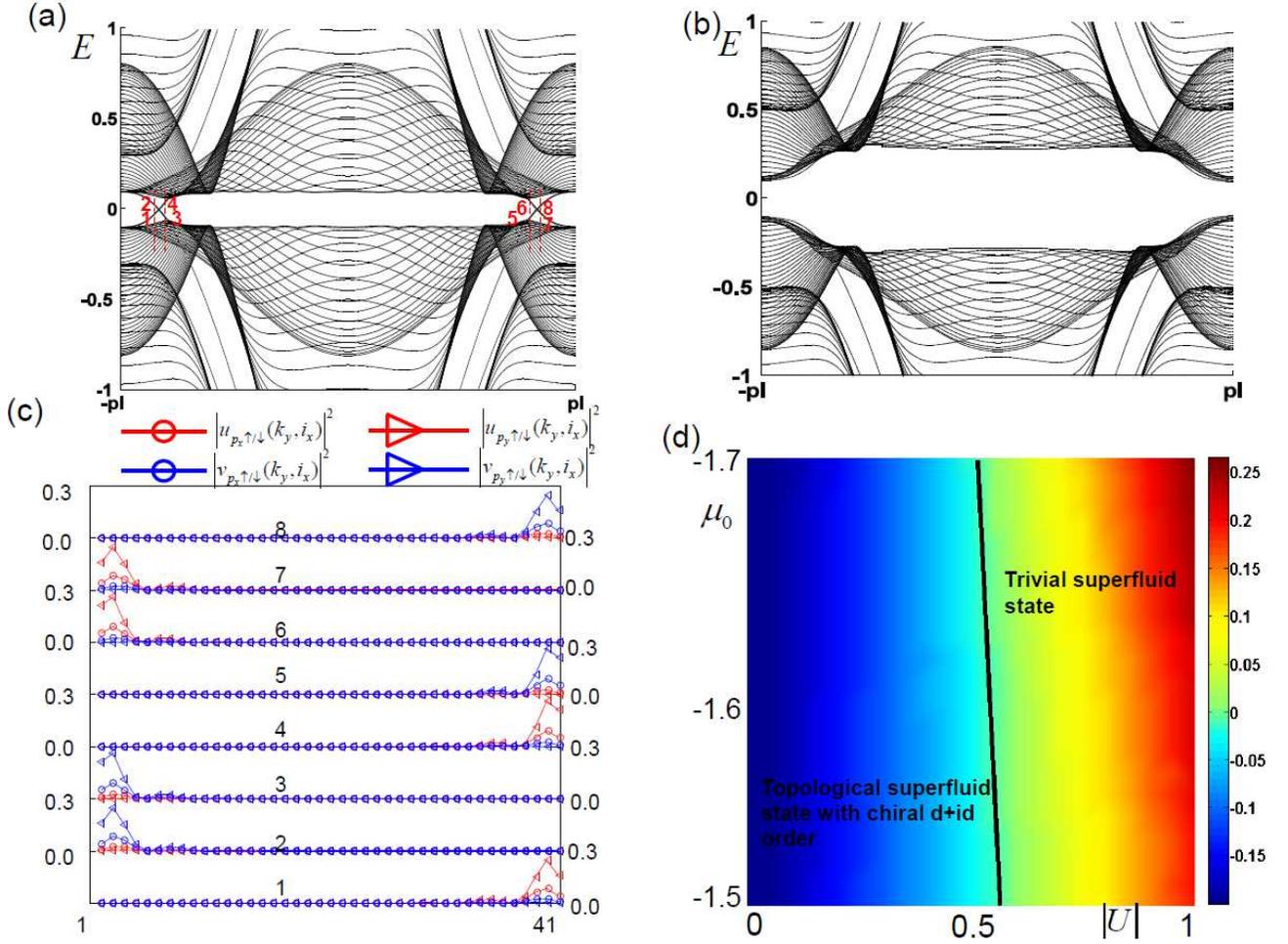}
\end{center}
\caption{(Color Online) (a) and (b) the edge spectra of the superfluid states
with $|U|=0.4$, $\mu_{0}=-1.6$ in (a) and $|U|=0.8$, $\mu_{0}=-1.6$ in (b).
The relevant $\Delta_{intra}=0.1$, $\Delta_{inter}=0.03i$ in (a) and
$\Delta_{intra}=0.3$, $\Delta_{inter}=0.08i$ in (b). Here, the y direction has
periodic boundary condition while the lattice number along x direction is set
to be $N_{x}=41$. (c) The amplitudes of wave-function of the in-gap states
labeled 1-8 in (a). Note that each point are double degeneracy by taking into
account the spin degree of freedom. Here, the red and blue \textquotedblleft
o\textquotedblright\ marks label particle-like $|u_{p_{x},\uparrow/\downarrow
}(k_{y},i_{x})|^{2}$and hole-like $|v_{p_{x},\uparrow/\downarrow}(k_{y}%
,i_{x})|^{2}$ while the red and blue \textquotedblleft$\vartriangleleft
$\textquotedblright\ marks label particle-like $|u_{p_{y},\uparrow/\downarrow
}(k_{y},i_{x})|^{2}$ and hole-like $|v_{p_{y},\uparrow/\downarrow}(k_{y}%
,i_{x})|^{2}$. (d) The phase diagram as change as chemical potential $\mu_{0}$
and interaction amplitude $|U|$. }%
\label{fig_top}%
\end{figure}%
\begin{equation}
W_{\sigma}={\oint}\frac{d\mathbf{k}}{2\pi}[\frac{F_{x}(k)}{F(k)}%
\mathbf{\nabla}\frac{F_{y}(k)}{F(k)}-(x\longleftrightarrow y)],
\label{winding number}%
\end{equation}
with $F(k)=\sqrt{F_{x}^{2}(k)+F_{y}^{2}(k)}$. The direct calculation gives
$W_{\sigma}=2$ in agreement with the pattern of the vector field
$[F_{x}(k),F_{y}(k)]$ as shown in Fig. \ref{fig_op}(b). Note that the total
winding number $W=W_{\uparrow}+W_{\downarrow}$ should be 4 when the spin
degree of freedom is taken into account. In the presence of $\Omega_{z}$, the
induced orbital hybridization lifts the degeneracy at M point. Then, the above
mapping does not work.

In the superfluid states, quasi-particle spectra are fully gapped and the
nonzero $\Omega_{z}$ breaks the pseudo-time-reversal symmetry. It is natural
to introduce the Chern number to characterize the topological properties of
the superfluid states. To show it, we consider the effective superfluid
Hamiltonian spanned in the effective Nambu basis: $\bar{\Psi}(k)=[\bar
{p}_{+,k,\uparrow},\bar{p}_{-,k,\uparrow},\bar{p}_{+,-k,\downarrow}^{\dag
},\bar{p}_{-,-k,\downarrow}^{\dag},\bar{p}_{+,k,\downarrow},\bar
{p}_{-,k,\downarrow},\bar{p}_{+,-k,\uparrow}^{\dag},\bar{p}_{-,-k,\uparrow
}^{\dag}]^{t}$,
\begin{equation}
\bar{H}_{mf}=\sum_{k}\bar{\Psi}^{\dag}(k)[\bar{H}_{tb}(k)+\bar{H}_{int}%
^{p}(k)]\bar{\Psi}(k), \label{H_effmf}%
\end{equation}
with
\begin{equation}
\bar{H}_{int}^{p}(k)=s_{z}\otimes\left[
\begin{array}
[c]{cc}
& \mathbf{\bar{\Delta}}\\
\mathbf{\bar{\Delta}}^{\dag} &
\end{array}
\right]  , \label{H_pair_eff}%
\end{equation}
and%
\[
\mathbf{\bar{\Delta}}=\left[
\begin{array}
[c]{cc}%
\left\vert \Delta_{inter}\right\vert  & \Delta_{intra}\\
\Delta_{intra} & -\left\vert \Delta_{inter}\right\vert
\end{array}
\right]  .
\]
Here $\Delta_{intra}=\Delta_{22}$ and $|\Delta_{inter}|=|\Delta_{23}|$. Upon
an unitary rotation\cite{Sato2009PRL}, we can obtain a dual form of the
Hamiltonian, i.e.,
\begin{equation}
\bar{H}_{mf}^{D}=S\bar{H}_{mf}S^{\dag}, \label{HmfD0}%
\end{equation}
where%
\begin{equation}
S=\frac{1}{\sqrt{2}}\left[
\begin{array}
[c]{cccc}%
1 & s_{x} &  & \\
is_{y} & -s_{z} &  & \\
&  & 1 & -s_{x}\\
&  & -is_{y} & -s_{z}%
\end{array}
\right]  , \label{Smatrix}%
\end{equation}

\[
\bar{H}_{mf}^{D}(k)=\left[
\begin{array}
[c]{cc}%
\bar{H}_{mf}^{D+}(k) & \\
& \bar{H}_{mf}^{D-}(k)
\end{array}
\right]
\]

\begin{align}
\bar{H}_{mf}^{D\pm}(k)  &  =\left[
\begin{array}
[c]{cc}%
\Delta_{intra}-h\Omega_{z}s_{z} & \pm h(k)\\
\pm h^{\dag}(k) & -\Delta_{intra}+h\Omega_{z}s_{z}%
\end{array}
\right]  ,\label{HmfD}\\
h(k)  &  =is_{y}[-\frac{\xi_{+}(k)}{2}+\mu_{0}+\frac{\xi_{-}(k)}{2}%
s_{x}\nonumber\\
&  -\xi_{xy}(k)s_{y}+is_{y}|\Delta_{inter}|]. \label{h4}%
\end{align}
In the dual Hamiltonian $\bar{H}_{mf}^{D\pm}(k)$ shown in Eq. (\ref{HmfD}),
[$\frac{\xi_{-}(k)}{2},\xi_{xy}(k)$] resembles two components of pairing order
parameters of the chiral $d$-wave superfluid and $\Delta_{inter}$ corresponds
to the mixed $s$-wave component. \textquotedblleft$\Delta_{intra}\pm
h\Omega_{z}$\textquotedblright\ is the pseudo-kinetic energy with
$k$-independent, and resembles kinetic energy term \textquotedblleft%
$\frac{k^{2}\pm k_{F}^{2}}{2m}$\textquotedblright\ of the chiral $d$-wave
superfluid when $\mu_{0}$ is set to satisfy the condition $\mu_{0}=\frac{1}%
{2}\xi_{+}(\pi,\pi)$. Then, the dual Hamiltonian $\bar{H}_{mf}^{D\pm}(k)$
resembles the standard Hamiltonian describing the chiral $d$-wave
superconductors\cite{Laughlin1998PRL,Volovik1997JETP}, and belongs to class
$C$ according to the classification by Schnyder \textit{et al}%
\cite{Schnyder2008PRB}. Here, $\Delta_{inter}$ by itself cannot drive the
gap-closing condition, because it is much smaller than $\Delta_{intra}$ and
Fermi energy. Therefore, the small $\Delta_{inter}$ can be absorbed and set to
zero. The topological nontrivial superfluid states can be achieved under the
condition\cite{Volovik1997JETP} $\Delta_{intra}<$ $h\Omega_{z}$ when $\mu
_{0}=\frac{1}{2}\xi_{+}(\pi,\pi)$, which naturally corresponds to the
weak-coupling condition$\frac{k^{2}-k_{F}^{2}}{2m}<0$\cite{Read2000PRB}.
For
the general case with arbitrary $\mu_{0}$, one can obtain nontrivial
superfluid states if $f(\Omega_{z},\mu_{0},\Delta_{intra})>0$ with $f(\Omega_{z},\mu_{0},\Delta_{intra})$ shown in Eq.(\ref{condition}), and trivial superfluid states if $f(\Omega_{z},\mu_{0},\Delta_{intra})<0$. The topological phase transition condition coincides with the gap-closing condition with
$f(\Omega_{z},\mu_{0},\Delta_{intra})=0$. The phase diagram separating the topological trivial and
non-trivial superfluid phases is plotted in Fig. \ref{fig_top}(d) according
to phase transition condition $f(\Omega_{z},\mu_{0},\Delta_{intra})=0$.
\begin{equation}
f(\Omega_{z},\mu_{0},\Delta_{intra})=|h\Omega_{z}|-\sqrt{\Delta_{intra}%
^{2}+\left[  \frac{\xi_{+}(\pi,\pi)}{2}-\mu_{0}\right]  ^{2}}.
\label{condition}%
\end{equation}
The nontrivial topological nature of the superfluid states can be
characterized by the Chern number,
\begin{equation}
\mathcal{C}_{s}=\frac{i}{2\pi}\sum_{E_{n}<0}%
%TCIMACRO{\dint \nolimits_{BZ}}%
%BeginExpansion
{\displaystyle\int\nolimits_{BZ}}
%EndExpansion
d\mathbf{k}\langle\nabla_{k}u_{s,n}(\mathbf{k})|\times|\nabla_{k}%
u_{s,n}(\mathbf{k})\rangle, \label{Chern_number}%
\end{equation}
with $u_{s,n}(\mathbf{k})$ the Bloch functions of occupied quasi-particle
states with $s=up$ and $down$ to label the the up-block and down-block parts
of Hamiltonian in Eq. (\ref{Hmf}). The straightforward calculations give
$\mathcal{C}_{up}=\mathcal{C}_{down}=2$ for $h\Omega_{z}>0$ and $\mathcal{C}%
_{up}=\mathcal{C}_{down}=-2$ for $h\Omega_{z}<0$ under the condition
$f(\Omega_{z},\mu_{0},\Delta_{intra})>0$, which means the inverse local
rotation corresponds to reverse chirality. From the bulk-edge correspondence,
the quasi-particle spectra have two chiral gapless edge states at the open
boundary shown in Fig. \ref{fig_top}(a) and no gapless edge states emerge in
trivial superfluid state shown in Fig.\ref{fig_top}(b). The local feature of
the edge states in the Fig. \ref{fig_top}(a) are explicitly demonstrated
through the amplitude distributions of the wave-functions shown in Fig.
\ref{fig_top}(c).

\section{Mass density modulation from the harmonic confining potential}

\begin{figure}[ptb]
\begin{center}
\includegraphics[width=1.0\linewidth]{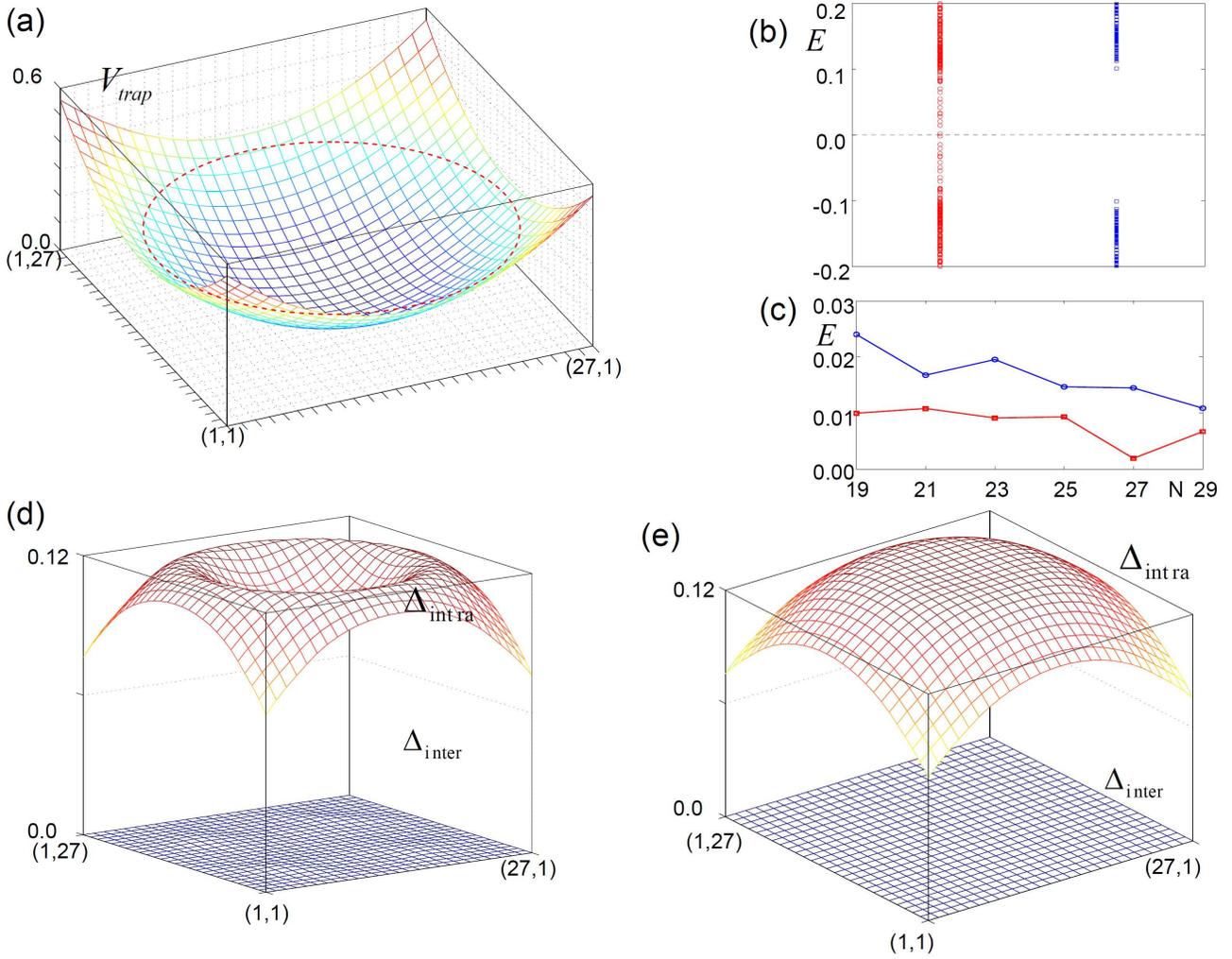}
\end{center}
\caption{(Color Online) (a) The pattern of weak harmonic confining potential
$V_{trap}$ with $V_{t}=1.2/N_{x}N_{y}$ in lattice space. The red-dashed circle
denotes a spacial barrier structure of the potential, which separates the two
different superfluid states. (b) The spectra of the superfluid states with the
red \textquotedblleft o\textquotedblright\ marks and the blue
\textquotedblleft$\square$\textquotedblright\ marks corresponding to the case
with $h\Omega_{z}=0.2$ and $h\Omega_{z}=0.05$ respectively. (c) The energy
levels of the in-gap fermion zero modes as function of lattice size N$\times
$N. Here, the red \textquotedblleft$\square$\textquotedblright\ marks and the
blue\textquotedblleft o\textquotedblright\ marks correspond to the first and
second lowest positive energy levels, and $h\Omega_{z}=0.2$. (d) and (e) The
distributions of superfluid order parameters including intra-orbital and
inter-orbital parts in lattice space with lattice size $(N_{x},N_{y})=(27,27)$
and $h\Omega_{z}=0.2$ in (c) and $\Omega_{z}=0.05$ in (d). Here, the
interaction strength $|U|=0.8$, chemical potential $\mu_{0}=-1.6$, and the
periodic boundary condition is applied. Other parameters are same as those in
Fig. 1. }%
\label{fig_poten}%
\end{figure}

Now, we consider the realistic case with nonzero harmonic confining potential
in Eq. (\ref{Vtrap}), and the pattern of $V_{trap}(i_{x},i_{y})$ is shown in
Fig. \ref{fig_poten}(a) with $V_{t}=1.2/N_{x}N_{y}$. We perform the
self-consistent calculations about the Bogoliubov-de Gennes (BdG) Hamiltonian
$H_{tb}+H_{int}^{p}$ in Eqs. (\ref{Ham_tb}) and (\ref{Hintmf}) in lattice
space. The quasi-particle spectra and the distribution of superfluid order
parameters are shown in Figs. \ref{fig_poten}(b), \ref{fig_poten}(d), and
\ref{fig_poten}(e) for two different $h\Omega_{z}=0.2$ and $h\Omega_{z}=0.05$
under the periodic boundary condition. We find that the amplitudes of
superfluid order parameters in both cases are similar from Fig.
\ref{fig_poten}(d) and \ref{fig_poten}(e), but the quasi-particle spectra are
quite different from Fig. \ref{fig_poten}(b) with in-gap fermion modes for
$h\Omega_{z}=0.2$ and without in-gap fermion modes for $h\Omega_{z}=0.05$. The
reason lies in that $V_{trap}(i_{x},i_{y})$ forms a spatial barrier structure
[The position is marked with red-dashed circle in Fig. \ref{fig_poten}(a)]
separating the nontrivial superfluid state with $f(\Omega_{z},\mu_{i}%
,\Delta_{intra})>0$ and trivial superfluid state with $f(\Omega_{z},\mu
_{i},\Delta_{intra})<0$ for $\Omega_{z}=0.2$.
Note that $\mu_{i}=\mu_{0}+V_{trap}(i_{x},i_{y})$, thus the position of spatial barrier coincides with the gap-closing condition with $f(\Omega_{z},\mu_{i},\Delta_{intra})=0$
. For fixed $\mu_{0}$ and $V_{t}$, one can find that $f(\Omega_{z},\mu
_{i},\Delta_{intra})$ is always smaller than zero when $\Omega_{z}=0.05$. The
superfluid is always trivial, because $\Omega_{z}=0.05$ is too small to
overcome the gap-closing condition $f(\Omega_{z},\mu_{i},\Delta_{intra})=0$.
The spatial barrier traps in-gap fermion modes and accumulates atoms when the
negative energy states are occupied\cite{Volovik1997JETP,Volovik2003book}. The
in-gap fermion modes trapped by the spatial barrier have the same origin as
the fermion modes in spectrum of the Caroli-de Gennes-Matricon bound states in
the vortex core\cite{Caroli1964PL}.

In the low-energy limit, the spectrum of in-gap fermion modes in terms of the
angular momentum $Q$ takes the following form under the axisymmetric
condition\cite{Volovik1997JETP,Volovik2003book},%

\begin{equation}
E_{a}(Q)=\omega_{a}(Q-Q_{a}), \label{E_boundstate}%
\end{equation}
where $\omega_{a}=c_{a}/R$ is the angular velocity of the rotation along the
spatial barrier with $R$ the radius of spatial barrier of $V_{trap}%
(i_{x},i_{y})$\cite{Note1}, $a$ labels the $a$th branch, and $Q_{a}=\hbar
k_{a}R$. The total number of the branches is four according to the index
theorem\cite{Volovik2003book} when the spin degree of freedom is taken into
account. In the absence of external driving, the energy of in-gap fermion
modes is $E_{a}(0)=-\omega_{a}Q_{a}$. In the square lattice space, the
circular rotation symmetry $SO(2)$ for Eq. (\ref{E_boundstate}) is broken down
to $C_{4}$ symmetry, and the Fermi velocity is strongly anisotropic and the
superfluid order parameters are highly inhomogeneous. $Q_{a}$ can only take
the discrete values under the constraint of $C_{4}$ symmetry. Correspondingly,
the energy levels of the in-gap fermion modes trapped by the spatial barrier
are discrete [see Fig. \ref{fig_poten}(b) for details], and several energy
levels close to zero usually correspond to in-gap fermion modes trapped by the
spatial barrier. \begin{figure}[ptb]
\begin{center}
\includegraphics[width=1.0\linewidth]{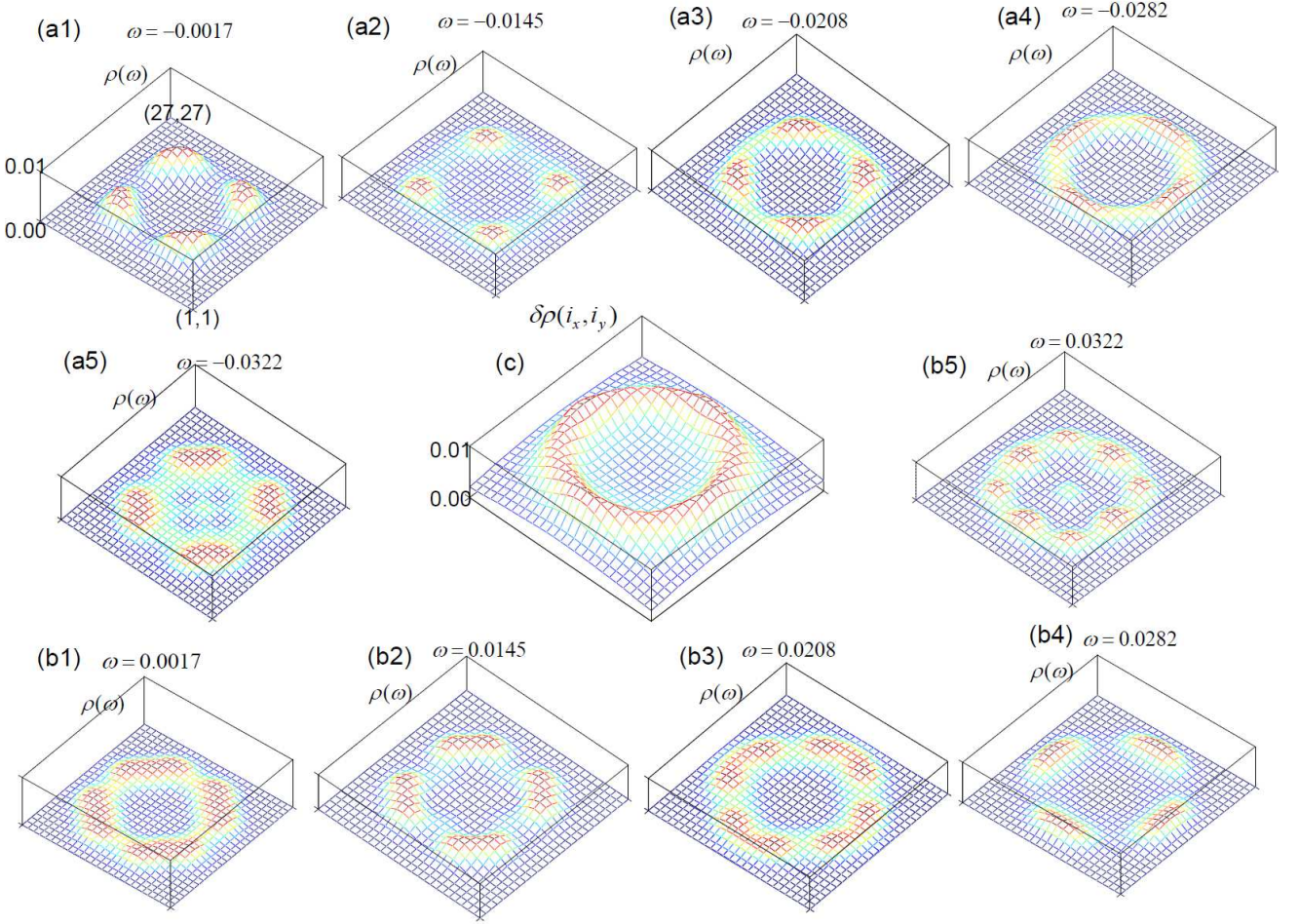}
\end{center}
\caption{(Color Online) (a1)-(a5) The distribution of LDOS defined in Eq.
(\ref{Den}) for five fermion modes with negative energy close to zero.
(b1)-(b5) The distribution of LDOS defined in Eq. (\ref{Den}) for five fermion
modes with positive energy close to zero. (c) The distribution of change of
LDOS between $h\Omega_{z}=0.2$ and $h\Omega_{z}=0.22$. Here, the parameters
are same as those in Fig. 4. }%
\label{fig_DOS}%
\end{figure}

The localization feature of the in-gap fermion modes trapped by the spatial
barrier can be reflected by the local density of states (LDOS), which is
calculated by%

\begin{equation}
\rho_{i}(\omega)=\sum_{n,l,\sigma}[|u_{i,l\sigma}^{n}|^{2}\delta(E_{n}%
-\omega)+|v_{i,l\sigma}^{n}|^{2}\delta(E_{n}+\omega)], \label{Den}%
\end{equation}
where $u_{i,l\sigma}^{n}$ and $v_{i,l\sigma}^{n}$ are the particle-like and
hole-like components of eigenstate with quasi-particle energy $E_{n}$ at site
$i$ and orbital $l$. The LDOS of the five in-gap fermion modes with the
highest negative energy are shown in Fig. \ref{fig_DOS}(a1)-(a5), from which
we can find that four levels with energy $-0.0017,-0.0145,-0.0208,-0.0282$ are
the fermion modes which are trapped by the spatial barrier. To make a
comparison, the level with energy $-0.0322$ is the extended state. We also
plot the LDOS of the the five levels with the lowest positive energy in Fig.
\ref{fig_DOS}(b1)-(b5) for comparison. Furthermore, we find that the highest
negative energy level and the lowest positive energy level approach zero
energy with increasing the lattice size N$\times$N [see Fig. \ref{fig_poten}%
(c) for details].

In the presence of external driving, the spectrum of the in-gap fermion modes
is a function of the angular momentum $Q$ from the external driving, and the
in-gap fermion modes could cross the zero energy and form the variation of the
mass current. The change of the mass current trapped in the spatial barrier
is\cite{Volovik1997JETP}%

\begin{equation}
\delta I_{M}=\frac{\hslash}{8\pi}\sum_{a}\delta(k_{a}^{2}), \label{current1}%
\end{equation}
where we have assumed the thickness along $z$ direction to be unity. The extra
1/2 in denominator is added to compensate the double count due to the
particle-hole symmetry. Generally, there are several external perturbations
which can be introduced to be the driving force to move the in-gap fermion
modes cross the zero energy, such as the modulations of $V_{1}$ and $V_{2}$ in
Eq. (\ref{potentialVxy}) to deform the $\frac{\xi_{-}(k)}{2}$ and $\xi
_{xy}(k)$ and introducing additional laser beam to modulation the trapping
potential. Here, we consider a more convenient method. From Eq.
(\ref{E_boundstate}), it is straightforward to inject non-zero $Q$ into the
superfluid state through slight modulation of local rotating frequency
$\Omega_{z}$. As a consequence, the in-gap fermion modes can be driven to
cross the zero energy by the non-zero $\delta\Omega_{z}$. If we further assume
that all the in-gap fermion modes trapped in the spatial barrier have the
relation $\frac{\hslash^{2}\delta(k_{a}^{2})}{2m}\sim h\delta\Omega_{z}$, we
can obtain that the response of change of mass current to the modulation of
the rotating frequency $\delta I_{M}$ $\sim\frac{m\delta\Omega_{z}}{2}\sum
_{a}sgn(c_{a})$ with the summation involving all the in-gap fermion modes
cross zero energy. However, in the square lattices, we can find that the
$k_{a}$ is different for different $a$-th branch from Fig. \ref{fig_DOS}. As a
good approximation, we can define an effective $\langle k\rangle$ to remove
the difference of different $k_{a}$, and $\langle k\rangle$ can be replaced
with the averge Fermi momentum $\langle k_{F}\rangle$. Then, we can obtain
that the modulation of mass current density is proportional to the change of
the LDOS, i.e.,%

\begin{equation}
\delta j_{M}(i_{x},i_{y})\varpropto\delta\rho(i_{x},i_{y}), \label{current2}%
\end{equation}
with%

\begin{align}
\delta\rho(i_{x},i_{y})  &  =\rho(i_{x},i_{y})|_{\Omega_{z}+\delta\Omega_{z}%
}-\rho(i_{x},i_{y})|_{\Omega_{z}},\label{current3}\\
\rho(i_{x},i_{y})|_{\Omega_{z}}  &  =\sum_{n,l,\sigma}|v_{l\sigma}^{n}%
(i_{x},i_{y})|^{2}\theta(-E_{n})|_{\Omega_{z}}. \label{current4}%
\end{align}
The pattern of $\delta\rho(i_{x},i_{y})$ for $\delta\Omega_{z}=0.02/h$ is
shown in Fig. \ref{fig_DOS}(c), from which we can find that the mass current
is trapped around the spatial barrier.

\section{Discussions and conclusions}

In terms of experiment, the fermion atoms can be selected as lithium $^{6}$Li,
two internal states can be selected as $^{2}$S$_{1/2}$ with M=$\pm\frac{1}{2}%
$. The principal fluorescence line from $^{2}$S$_{1/2}$ to $^{2}$P is at 670.8
nm. Therefore, a Nd:YAG-laser with 532 nm could be selected to be the light
source to realize the optical potential with the lattice constant $a=$ 532 nm.
The recoil energy $E_{R}\sim h\times100$ KHz. The local rotation around each
potential minimum has been experimentally realized through inserting
electrooptic phase modulators into the beams forming the 2D lattice potential,
and the relevant rotating frequency $\Omega_{z}$ can be turned with large
flexibility\cite{GemelkeARXIV2010}. From the energy bands in Fig.
\ref{fig_laser}, we can estimate that it is enough for $\Omega_{z}\sim
h\times2$ KHz to satisfy the topological superfluid condition.

 In the presence of the harmonic trap, it has been shown that the
local density approximation(LDA) breaks down for trapped non-interacting
bosons in p-orbital bands, and increasing the interactions and optical lattice
potentials can suppress anisotropy of condensate density\cite{Pinheiro2012PRA}%
. However, the picture is different for trapped non-interacting fermions in
p-orbital bands due to the different statistics. It is shown that the
hard-core boson known as Tonks-Girardeau boson with infinitely repulsive
interactions can be mapped into non-interacting free fermion in one
dimensional
limit\cite{Girardeau1960JMP,Paredes2004Nature,Kinoshita2004Science}. Thus, the
boson with infinitely repulsive interactions is roughly equivalent to free
fermion even in two dimensional system. Such effective \textquotedblleft repulsive
interactions\textquotedblright  can suppress the anisotropy of condensate density, and guarantee
the validity of LDA in system with trapped fermions in p-orbital bands.
Furthermore, the tunability of the optical lattice potential and quite small
trap potential can further reduce the anisotropy of condensate density. Though the breaking down of LDA can be suppressed, the particle density per site will inevitably vary and the s-orbital atoms will thereby shift the onsite energies for p-orbital atoms in the presence of the trap. Thanks to the small trap potential, one can expect that the density fluctuations of the both trapped s-orbital and p-orbital atoms should be small, and the main results throughout the paper are not changed qualitatively.

The change of the mass current and the accumulation of the atoms around the
spatial barrier can be spatially resolved with the radio-frequency
spectroscopy\cite{Gupta2003Science,Regal2003PRL,Shin2007PRL}. Besides the
radio-frequency spectroscopy, the recently developed matter-wave interference
technique\cite{Kock2015PRL} is a more powerful tool, which can directly
represent the phase properties of the superfluid order parameter. More
remarkably, one can reconstruct the spatial geometry of certain low-energy
in-gap fermion modes and verify the formation of the spatial barrier
structure, both of which are the key signatures in our
proposal.

In summary, we propose that the superfluid states of fermions with a chiral
$d$-wave order can be implemented in a rotating optical lattice where the
orbital degrees of freedom play a key role. Our proposal presents an
alternative route to realize the topological superfluids with chiral
even-frequency order in the absence of the spin-orbital coupling. Furthermore,
we show that the intrinsic harmonic confining potential can form a circular
spatial barrier structure which accumulates atoms and support a mass current
under the injection of small angular momentum as driving force. The mass
current associated with the accumulated atoms can be experimentally detected,
and provides a signature to verify the emergence of topological superfluid
state with chiral $d$-wave order in a rotating optical lattice.

\begin{acknowledgments}
We thank S.-Q. Shen, S. Z. Zhang, D. W. Zhang, G. C. Liu for useful discussions. This work
was supported by NSFC under Grants No. 11674331, No. 11274032, No.11625415, the Ministry of Science and Technology of China (Grant number: 2017YFA0303200), and by 100 Talents Programme
of CAS.
\end{acknowledgments}

\end{document}